# Mobile-service based Max-Min Fairness Resource Scheduling for Heterogeneous Vehicular Networks

**Yu Zhang[1], Ke Xiong*[2,3], Fengping An[1], Xiaofei DI [2], Jingtao SU[1]**

1. School of Computer and Communication Engineering, University of Science and Technology Beijing, Beijing 100083, P. R. China
2. School of Computer and Information Technology, Beijing Jiaotong University, Beijing 100044, P. R. China
3. National Mobile Communications Research Laboratory, Southeast University, Nanjing 210018, P. R. China

**Abstract:** This paper investigates the resource scheduling for heterogeneous vehicular networks, where some moving vehicles are selected and scheduled as helping relays to assist information transmission between the roadside infrastructure and other moving vehicles. For such a system, we propose a mobile-service based max-min fairness resource scheduling scheme, where service amount which is more suitable for high mobility scenarios is adopted to characterize the information transmission capacity of the links and the max-min criteria is adopted to meet the fairness requirement of the moving vehicles. Simulation results demonstrate the effectiveness of our proposed scheme. It is shown that our proposed scheme archives higher throughput and better fairness compared with random scheduling scheme and non relaying scheme.

## I. INTRODUCTION

Conventionally, vehicular networks in short range were supported by vehicle-to-vehicle (V2V) links, involving smart vehicles equipped with on-board computers and sensors (e.g., radar, ladar, etc.) and multiple network interface protocols (e.g., IEEE 802.11p, Bluetooth, etc.). The dedicated short-range communication (DSRC) or wireless accessing vehicular environment (WAVE) multi-hop paradigm was used for V2V transmission and exploits the flooding of information of vehicular data applications [1]. However, connectivity disruptions in V2V transmission often occurs because of quick topology network changes, vehicle speed, especially when vehicles are in sparse (i.e., low density) or totally disconnected scenarios. Another transmission mode of vehicular networks is to use the longer-range vehicular connectivity provided by pre-existing network infrastructure like wireless access is proposed, in which heterogeneous emerging wireless technologies, such as 3G cellular systems, long-term evolution (LTE), IEEE 802.11 and IEEE 802.16e, are integrated for effective vehicle-to-infrastructure (V2I) transmissions [2]. With the help of V2I transmission, some Internet-based information services can be provided for vehicular networks. However, the limitation of V2I links is that when vehicle moves relatively far away from the roadside infrastructure, the transmission capability over the V2I link will suffer a great degradation.

To inherit the advantages of both V2V and V2I links, the heterogeneous vehicular network (HetVeNet) architecture was proposed, where cooperative communication [3,4] is introduced to help the information transmission between the roadside infrastructure and the moving vehicle. That is, some moving vehicles are selected as helping relays to extend the limited communication range of V2I links by V2V links. By such a heterogeneous networking, the information transmission of vehicular networks can be greatly enhanced. Therefore, great deals of information applications are expected to be realized in future heterogeneous vehicular networks.

So far, many works have been done on designing efficient HetVeNet. An advanced vehicular relaying technique for enhanced connectivity in densely populated urban areas was proposed in [5]. A cooperative traffic transmission algorithm in a joint vehicular Ad Hoc networks and LTE hybrid network architecture was proposed in [6]. In [7], weighted sum rate was maximized by scheduling the V2V and V2I links, where bipartite graph method was proposed. In [8], it proposed a two-dimensional-multi-choice knapsack problem based scheduling scheme to achieve higher system spectral efficiency. In [9], it investigated the cooperative spectrum allocation with QoS support in cognitive cooperative vehicular ad hoc networks. Moreover, a novel heterogeneous vehicular network protocol based switching decision was proposed in [10] and a channel equalization method was proposed in [11] to reduce the demand of bit length for cooperative vehicular networks.

Besides, one of the most important works in designing efficient HetVeNet is the resource scheduling. For example, in vehicular networks, there are many moving vehicles, which ones should be directly connected with the roadside infrastructure via V2I links and which ones should be connected with the roadside infrastructure with the help of V2V links may great impact the system performance. Although some works began to investigated the resource scheduling for HetVeNet, see e.g., [7,8], there are still some problems requiring to be studied. One of the problems is that in most existing works, they scheduled the V2I and V2V links of heterogeneous vehicular network on the basis of instantaneous achievable information rate (AIR) of the links. It is known that the accurate calculation on instantaneous rate relies on the accurate channel state information (CSI) of links. In practical systems, CSI and instantaneous rate are evaluated periodically, which means the CSI obtained at time $t$ is also used for time $t+\Delta t$ for $\Delta t$. In low mobility scenarios, the CSI of time $t$ is very similar to that of time $t+\Delta t$. Thus, the ARI-based resource scheduling can approach the system capacity upper bound. But, in high mobility scenarios, due to the high moving speed, the CSI of time $t$ may be very difference from that of time $t+\Delta t$. In this case, if we still use the AIR-based resource scheduling, to get more precise result, the time period of $\Delta t$ should be reduced, which means that more times of CSI calculating and system scheduling should be performed, resulting much heavier calculation and scheduling complexity.

Thus, the goal of this work is to design a cooperative relaying scheduling with low complexity and high accuracy adapting to high mobility for HetVeNet. To this end, we propose a cooperative heterogeneous vehicular network fairness resource scheduling scheme called mobile-service based max-min fairness resource scheduling (MS- MAXMIN), which benefits from the forecast of short-term mobile service amount and the fairness among the moving vehicles is fully considered. In the proposed scheme, it predicts the short-term mobile service amount to characterize the information transmission capacity of the links. Compared with AIR, it can characterize the information transmission capability more accurate in mobility scenarios. Based on the predicted short-term mobile service amount, it use the max-min mobile service amount rule to allocate communication resources fairly. The key features of the proposed scheme are as follows.

1) MS-MAXMIN is based on the predicting of short-term mobile service amount of V2I and V2V links in HetVeNet, as the knowledge of vehicles including location, speed and direction in practical system can be obtained by using some effective methods, so it does not need to rapidly and frequently exchange neighborhood information.

2) MS-MAXMIN takes full consideration of fairness among moving vehicles, where the max-min fairness criteria is employed and the Jain's fairness index is used to evaluate its fairness behavior.

3) MS-MAXMIN is performed with a centralized manner to allocate communication resources in HetVeNet, where is assumed that a supercomputer or a cloud with supercomputing capability is employed in the system to perform the computing and scheduling. Simulation results show that our proposed scheme archives higher throughput and fairness compared with random scheduling scheme and traditional non-relay scheme.

The remainder of this paper is structured as follows. Section II describes the system model. Section III introduces the concept of short-term mobile service amount. In Section IV, the MS-MAXMIN scheme is proposed, and in Section V, extensive simulation results are provided. Finally some conclusions of the paper are summarized in Section VI.

## II. SYSTEM MODEL

### 2.1 Network Model

Consider a vehicular communication system as shown in Figure 1, where $N$ moving vehicular nodes (VN) on the road desire to communicate with a common infrastructure located on the roadside. Each VN can communicate with the infrastructure via direct V2I links between the infrastructure and VN in LTE-A networks [12]. Due to large propagation path loss, only the vehicles near to the infrastructure are able to acquire good communication service and the ones relatively far away from the infrastructure often experience poor communication service. In this case, the far vehicles (FV) may ask the ones with relatively high V2I transmission capacity as relay vehicles (RV) to help them forward information to/from the infrastructure by DSRC [13] V2V links.

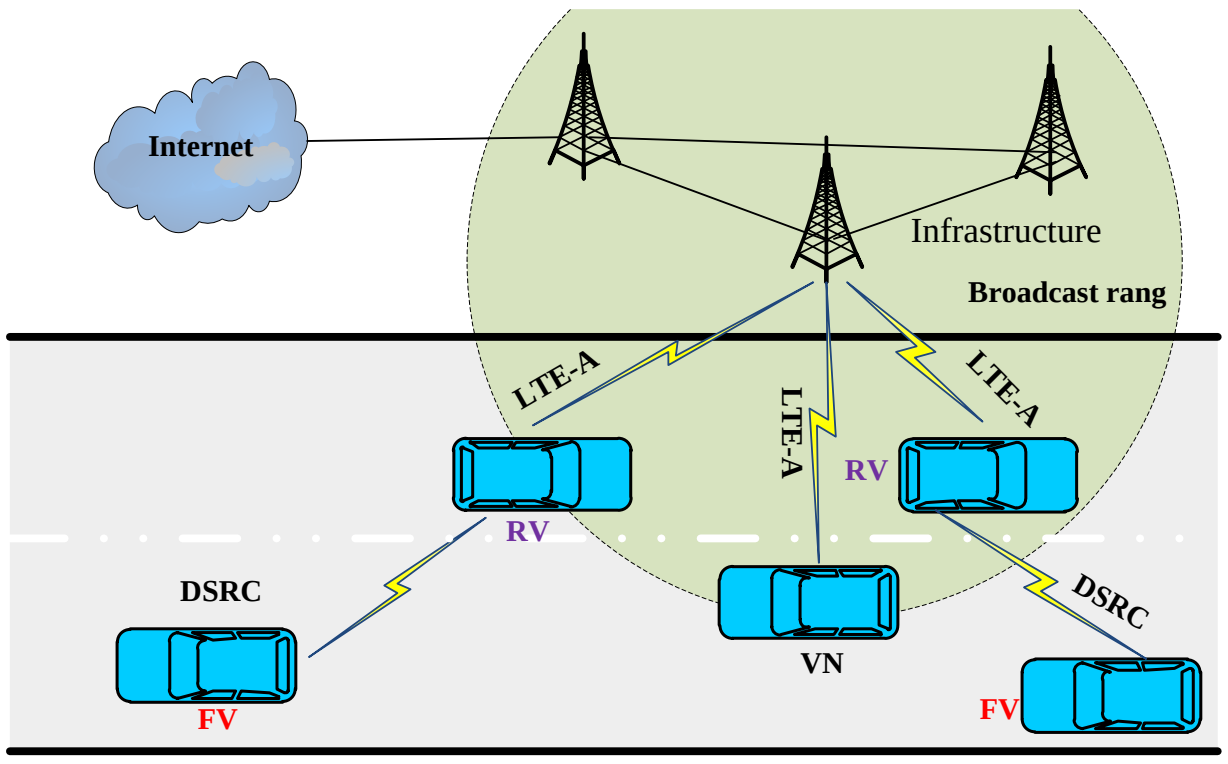

Fig. 1. HetVeNet model

### 2.2 Instantaneous Achievable Information Rate

In vehicular networks, the effect of large-scale fading (propagation path loss) generally is much more serious than that of small-scale fading. Similar to existing work on vehicular systems, see e.g., [14], we ignore the channel variation caused by small-scale fading and assume that the change of received signal strength only depends on the position shifts of vehicles. In this case, the path loss can be expressed as a function of the distance between the transmitter and the receiver,

$$LP(\mathrm{d}(t)) = \mathrm{F} + 10\alpha \log_{10}\left(\frac{\mathrm{d}(t)}{\mathrm{d}_0}\right), \quad (1)$$

where F is the path-loss attenuation at reference distance $\mathrm{d}_0$, which is affected by the carrier frequency, the heights of the transmitter and receiver antennas, different climate or geology conditions, etc. $\alpha$ is the path-loss exponent (usually $\alpha \geqslant 2$) and $\mathrm{d}(t)$ is the distance between the transmitter and the receiver at time $t$ [15]. As $LP(\mathrm{d(t)})$ also can be expressed by

$$LP(\mathrm{d(t)}) = 10\log_{10}\left(\frac{P_s}{P_r}\right). \quad (2)$$

According to (1) and (2), one can calculate the received power $P_r$ for a given transmit power $P_s$, as

$$P_r = \frac{P_s}{10^{\frac{LP(\mathrm{d}(t))}{10}}}. \quad (3)$$

Assuming that all links in the system are additive white Gaussian noise (AWGN) channel with a reference noise power $\sigma_N^2$. Therefore, at the moment $t$, the AIR between a transmit and its receiver is

$$C(t) = \log_2\left(1 + \frac{P_r}{\sigma_N^2}\right) = \log_2\left(1 + \frac{P_s}{\sigma_N^2 10^{\frac{LP(\mathrm{d}(t))}{10}}}\right), \quad (4)$$

where $\frac{P_s}{\sigma_N^2 10^{\frac{LP(\mathrm{d}(t))}{10}}}$ is the received signal to noise ratio at receiver.

For V2I link, LTE-A is adopted. F=128.1, $\mathrm{d}_0$=1000 and $\alpha$=3.76 for LTE-A in (1). So the path loss between the $i$-th VN ($i$=1,2,…,$N$) and the infrastructure can be expressed as

$$LP_{i\mathrm{I}}\left(\mathrm{d}_{i\mathrm{I}}^{\mathrm{V2I}}(t)\right) = 128.1 + 37.6\log_{10}\left(\frac{\mathrm{d}_{i\mathrm{I}}(t)}{1000}\right). \quad (5)$$

Global Positioning System (GPS) devices are often used to determine the location of vehicles in VANET [16]. Let $(\mathrm{x}_i(t), \mathrm{y}_i(t))$ denote the location of $i$-th VN at time $t$, $(\mathrm{x}_\mathrm{I}, \mathrm{y}_\mathrm{I})$ denote the location of infrastructure. Then the distance between the $i$-th VN and the infrastructure can be expressed as

$$\mathrm{d}_{i\mathrm{I}}(t) = \sqrt{(\mathrm{x}_i(t) - \mathrm{x}_\mathrm{I})^2 + (\mathrm{y}_i(t) - \mathrm{y}_\mathrm{I})^2}. \quad (6)$$

The entire LTE-A radio resources are divided into $N_{\mathrm{LTE}}$ resource blocks (RBs) along the time and/or frequency domain. A fair share of all the RBs among all $N$ VNs is adopted, which means each VN has $N_{\text{LTE-RB}} = \lfloor N_{\mathrm{LTE}} / N \rfloor$ RBs. So the AIR of V2I link is

$$C_{i\mathrm{I}}(t) = N_{\text{LTE-RB}} \log_2\left(1 + \frac{P_s}{\sigma_N^2 10^{\frac{LP_{i\mathrm{I}}(\mathrm{d}_{i\mathrm{I}}(t))}{10}}}\right). \quad (7)$$

For V2V link, DSRC is adopted. F=43.9, $\mathrm{d}_0$=1 and $\alpha$=2.75 for DSRC in (1). So the path loss between the $i$-th VN and the $j$-th VN can be expressed as

$$LP_{ij}\left(\mathrm{d}_{ij}(t)\right) = 43.9 + 27.5\log_{10}\left(\mathrm{d}_{ij}(t)\right). \quad (8)$$

Then the distance between the $i$-th VN and the $j$-th VN can be expressed as

$$\mathrm{d}_{ij}(t) = \sqrt{(\mathrm{x}_i(t) - \mathrm{x}_j(t))^2 + (\mathrm{y}_i(t) - \mathrm{y}_j(t))^2}. \quad (9)$$

The entire DSRC radio resources are also divided into $N_{\mathrm{DSRC}}$ resource blocks (RBs) along the time and/or frequency domain. A fair share of all the RBs among all $N_\mathrm{F}$ FVs is adopted, which means each FV has $N_{\text{DSRC-RB}} = \left\lfloor \frac{N_{\mathrm{DSRC}}}{N_\mathrm{F}} \right\rfloor$ RBs. So the AIR of V2V link is

$$C_{ij}(t) = N_{\text{DSRC-RB}} \log_2\left(1 + \frac{P_s}{\sigma_N^2 10^{\frac{LP_{ij}(\mathrm{d}_{ij}(t))}{10}}}\right). \quad (10)$$

## 2.3 Short-term Mobile Service Amount

Mobile service amount is defined as the integral of the instantaneous AIR over a given short time period $T$, similar as in [17, 18].

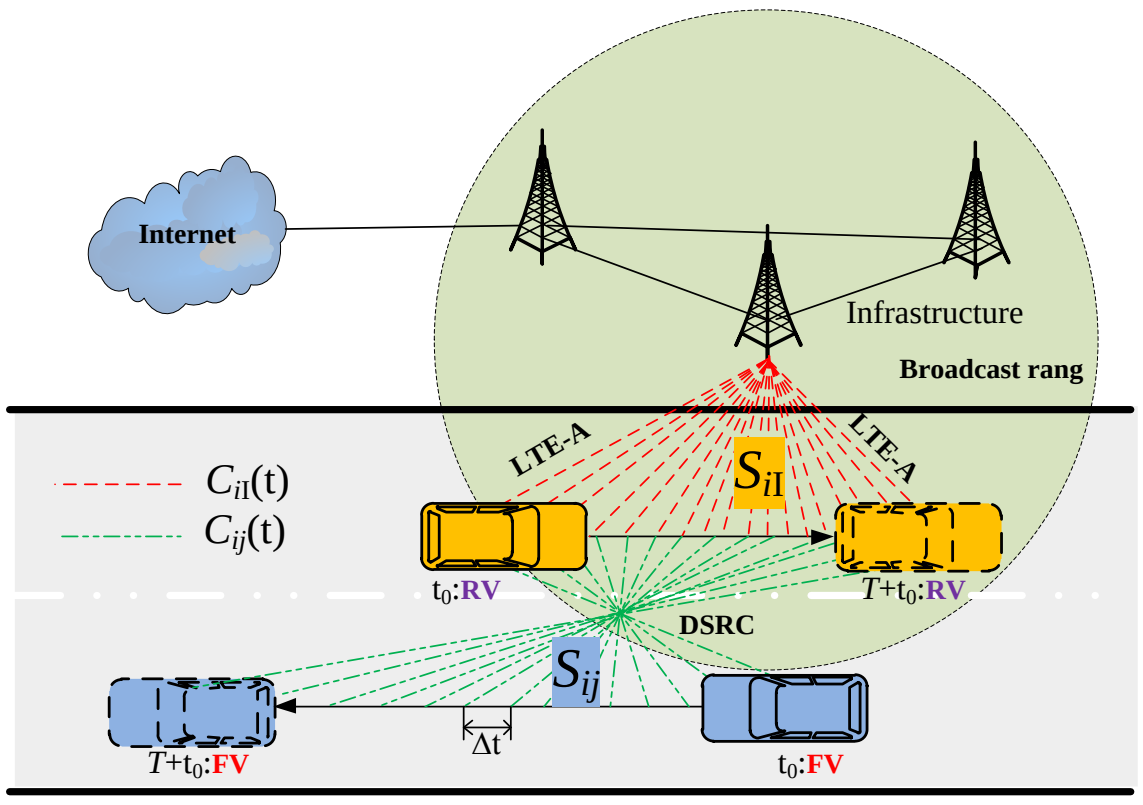

**Fig. 2. Mobile service amount of V2I and V2V links**

Specifically, in HetVeNet, mobile service amount of V2I and V2V as shown in Figure 2 is defined as $S_{i\mathrm{I}}$ and $S_{ij}$ $(1 \le i, j \le N)$, respectively:

$$\begin{aligned} S_{i\mathrm{I}} &= \int_{t_0}^{T+t_0} C_{i\mathrm{I}}(t)\,\mathrm{d}t \\ &= \int_{t_0}^{T+t_0} \log_2\left(1 + \frac{P_s}{\sigma_N^2 10^{\frac{LP_{i\mathrm{I}}(\mathrm{d}_{i\mathrm{I}}(t))}{10}}}\right)\mathrm{d}t \end{aligned} \quad (11)$$

and

$$S_{ij} = \int_{t_0}^{T+t_0} C_{ij}(t)\,\mathrm{d}t$$
$$= \int_{t_0}^{T+t_0} \log_2\left(1+\frac{P_s}{\sigma_N^2 10^{\frac{LP_{ij}(\mathrm{d}_{ij}(t))}{10}}}\right)\mathrm{d}t \quad , \quad (12)$$

where $i$-th VN is a RV, $j$-th VN is a FV and $t_0$ denotes the present moment. Parameters in $S_{i\mathrm{I}}$ and $S_{ij}$ are knowable at time $t_0$ except relative distance $\mathrm{d}_{\mathrm{V2I}}(t)$ and $\mathrm{d}_{\mathrm{V2V}}(t)$. Fortunately, since the movement of the vehicle is limited by the lane and the moving speed and direction of the vehicle can be regarded as a constant within a short term $T$. So the position of vehicle in a short term $T$ can be predicted by the position, speed and direction at time $t_0$. Then the $\mathrm{d}_{i\mathrm{I}}(t)$ and $\mathrm{d}_{ij}(t)$, and the $S_{i\mathrm{I}}$ and $S_{ij}$ can be evaluated in order.

### 2.4 Resource allocation based on Max-Min Fairness

While fairness in 802.11 networks has been the subject of a considerable body of literature, a large part of this literature is concerned with unfairness behavior in 802.11 networks (see e.g., [19], [20]).

As max–min fairness is an important kind of utility fairness [18], we study in this paper the resource allocation problem with the objective of maximizing the mobile service amount of the request that has the minimal mobile service amount (the max–min fairness in terms of mobile service amount).

The definition of max-min fairness [20]: A vector of $x \in R$ ($R$ is log-convex set) is max-min fair if and only if for every $y \in R$ if $y_i > x_i$ (for some component), then $y_j < x_j$ for some $j$ such that $x_j \le x_i$ .

It has been established that the 802.11 WLAN rate region is log-convex [21], so we can use the max-min fair analysis for VANET in this paper.

## III. MOBILE-SERVICE BASED MAX-MIN FAIRNESS RESOURCE SCHEDULING

In this section, we describe our proposed MS-MAXMIN. For this we first present some assumptions and then give the detail information of MS-MAXMIN.

### 3.1 Problem formulation

In HetVeNet, one RV help one FV to relay data toward infrastructure, which then forward received data to the remote server via wired networks.

All $N$ VNs fairly share all the $N_{\mathrm{LTE}}$ LTE-A RBs, and all $N_{\mathrm{F}}$ FVs fairly share all the $N_{\mathrm{DSRC}}$ DSRC RBs. FV gives its LTE-A RBs to the RV who helps to relay the FV's data, so that relaying the FV's data does not affect the RV to transmit its own data.

According (7), (10), (11), and (12), the AIR and the mobile service amount of $\mathrm{RV}_i$ is $C_i=C_{i\mathrm{I}}$ and $S_i=S_{i\mathrm{I}}$ respectively. Similarity, the AIR and the mobile service amount of $\mathrm{FV}_j$ is $C_j=\min(C_{i\mathrm{I}}, C_{ij})$ and $S_j=\min(S_{i\mathrm{I}}, S_{ij})$ respectively, when $\mathrm{RV}_i$ helps to relay the $\mathrm{FV}_j$'s data. Let $\mathrm{A}_{R_k F_k}$ represent a sorted set $R_k$ (elements are RVs) assigned to a sorted set $F_k$ (elements are FVs), in proper sequence ($r_i$ is assigned to $f_i$, $r_i$ is the $i$-th element in sorted set $R_k$ and $f_i$ is the $i$-th element in sorted set $F_k$). Note that the $r_i$ doesn't mean the $\mathrm{RV}_i$, and the $f_i$ doesn't mean the $\mathrm{FV}_i$. For example, a HetVeNet including four VNs is divided to a sorted set $\mathrm{R}_1$={$\mathrm{RV}_3$, $\mathrm{RV}_1$} and a sorted set $\mathrm{F}_1$={$\mathrm{FV}_4$, $\mathrm{FV}_2$}. In this case, $r_1$ is $\mathrm{RV}_3$, $r_2$ is $\mathrm{RV}_1$, $f_1$ is $\mathrm{FV}_4$ and $f_2$ is $\mathrm{FV}_2$. The problem of guaranteeing max-min fairness among the members of HetVeNet can be formulated by adopting the parameter $\mathrm{A}_{R_k F_k}$ ($0<k<N/2+1$):

$$\begin{aligned}
&\max_{\mathrm{A}_{R_k F_k}} \ \min_i (S_i) \\
&\text{s.t.} \quad \|F_k\| = N_{\mathrm{F}}, \ \|R_k\| + \|F_k\| = N \\
&\qquad S_{r_i \mathrm{I}} \ge S_{f_j \mathrm{I}}, \ \forall r_i \in R_k, \ f_j \in F_k \\
&\qquad S_{r_i \mathrm{I}} > 0, \ S_{f_j \mathrm{I}} > 0, \ S_{r_i f_j} > 0
\end{aligned}$$

### 3.2 Proposed MS-MAXMIN

In this paper, we focus on introduce mobile service amount into HetVeNet and apply max–min fairness to allocate communication resources among HetVeNet members.

The basic idea of our proposed scheme is as follows. Firstly, it computes the mobile service amount of all V2I and V2V links in HetVeNet according to the location, speed and direction of vehicles, and then it selects RV who has the high V2I mobile service amount. Finally, it matches FV to RV by the max–min fairness rule. The detailed process of he proposed MS-MAXMIN is shown in **Algorithm 1**.

| **Algorithm 1. Mobile-service based max-min fairness resource scheduling** | |
|---|---|
| **Step 1:** | Initialize the location, speed and direction of vehicles in HetVeNet at time $t$=$t_0$. Initialize the mobile service amount of V2I and V2V as $S_{i\mathrm{I}}$=0 and $S_{ij}$=0. |
| **Step 2:** | Compute the relative distance $\mathrm{d}_{i\mathrm{I}}(t)$ and $\mathrm{d}_{ij}(t)$ by (6) and (9), then calculate AIR $C_{i\mathrm{I}}(t)$ and $C_{ij}(t)$ in (7) and (10) for each V2I and V2V link. Update the mobile service amount of V2I and V2V as $S_{i\mathrm{I}}=S_{i\mathrm{I}}+\Delta t * C_{i\mathrm{I}}(t)$ and $S_{ij}=S_{ij}+\Delta t * C_{ij}(t)$, where $\Delta t=T/M$ is a small time interval. |
| **Step 3:** | Update $t$=$t$+$\Delta t$, calculate the location of vehicles in VANET at time $t$ according the speed and direction of vehicles and the previous location. If $t$ is not equal to $T$+$t_0$, back to step 2. If $t$ is equal to $T$+$t_0$, we get the mobile service amount vector of V2I as $S_{\text{V2I}}=(S_{1\text{I}},S_{2\text{I}},\cdots,S_{N\text{I}})_{1\times N}$ and the mobile service amount matrix of V2V as (13) then go to step 4. |
| **Step 4:** | Initialize the number of FV as $N_F$=0. |
| **Step 5:** | According $S_{\text{V2I}}$, the FV set $\aleph_{N_F}$ is composed of $N_F$ VNs that has the smallest V2I mobile service amount, and the RV set $\Re_{N_F}$ is composed of the other $N-N_F$ VNs. Let $\Im_{N_F}=\varnothing$. |
| **Step 6:** | If $\aleph_{N_F}=\varnothing$ go to step 9, else go to step 7. |
| **Step 7:** | Find the biggest $S_{ij}$ according $S_{\text{V2V}}$, $\text{VN}_i\in\Re_{N_F}$ and $\text{VN}_j\in\aleph_{N_F}$. |
| **Step 8:** | Add the selected $S_{ij}$ to $\Im_{N_F}$, delete the $\text{VN}_i$ from $\Re_{N_F}$ and $\text{VN}_j$ from $\aleph_{N_F}$, then back to step 6. |
| **Step 9:** | According the $S_{ij}$ in $\Im_{N_F}$, denote the sorted set $R_{N_F}$ and sorted set $F_{N_F}$. Let $m_{N_F}=\min_{S_{ij}\in\Im_{N_F}} S_{ij}$. Update $N_F$=$N_F$+1. If $N_F$>$N$/2, go to step 10, else back to step 5. |
| **Step 10:** | Select the smallest $m_{N_F}$ from the acquired $N$/2+1 channel resource allocation schemes as the MSFRS. |

$$S_{\text{V2V}}=\begin{matrix} & \text{VN}_1 & \text{VN}_2 & \text{VN}_3 & \cdots & \text{VN}_N \\ \text{VN}_1 & 0 & S_{1,2} & S_{1,3} & \cdots & S_{1,N} \\ \text{VN}_2 & S_{2,1} & 0 & S_{2,3} & \cdots & S_{2,N} \\ \text{VN}_3 & S_{3,1} & S_{3,2} & 0 & \cdots & S_{3,N} \\ \vdots & \vdots & \vdots & \vdots & \ddots & \vdots \\ \text{VN}_N & S_{N,1} & S_{N,2} & S_{N,3} & \cdots & S_{N,N} \end{matrix}_{N\times N} \tag{13}$$

## IV. SIMULATION RESULTS

In this section, we present some simulation results to discuss the performance of our proposed MS-MAXMIN in terms of total throughput of FVs, minimum VN rate and fairness. For comparison, six scheduling schemes are considered, i.e., our proposed mobile-service based max-min fairness resource scheduling (MS-MAXMIN), AIR based max-min fairness resource scheduling scheme (AR-MAXMIN), mobile-service based max-sum resource scheduling scheme (MS-MAXSUM), AIR based max-sum resource scheduling scheme (AR-MAXSUM), stochastic scheduling scheme (random method) and no relay scheme (no relay method).

We simulate the scenario as shown in Figure 1, where the vehicles straightly move on the road towards two opposing directions. An infrastructure is placed on the roadside with the distance of 15m between it and the road, whose position is also considered as the reference point. Each vehicle is represented by a vector ($x$,$y$,$v$), where $x$ and $y$ are the coordinates of $x$−axis and $y$−axis, respectively, and $v$ is the moving speed of the vehicle. The plus and minus value of $v$ denote the two directions. ($x$,$y$,$v$) is randomly generated with the limitation of the road area and the vehicle's speed setting. The coverage radius of the infrastructure is set as 1.5km and the absolute value of vehicle's moving speed is limited within 35m/s. These configurations will not change in the sequel unless otherwise specified.

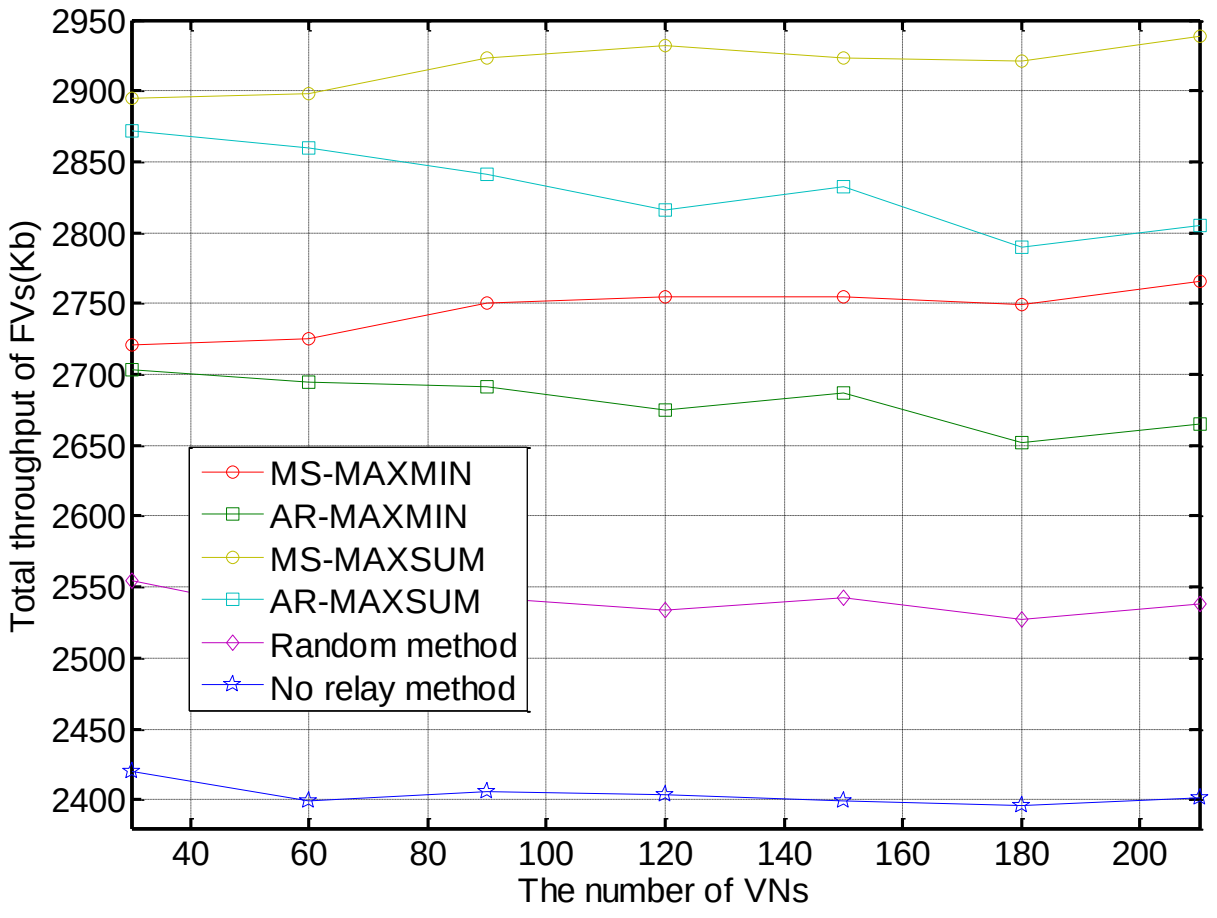

**Fig.3 Total throughput of FVs vs. the number of VNs.**

In Fig. 3 it can be observed that the gap between the schemes based on mobile-service (MS-MAXMIN, MS-MAXSUM) and the schemes based on AIR (AR-MAXMIN, AR-MAXSUM) in the term of total FVs throughput grows with the increase in the size of HetVeNet. It shows that mobile-service is a better target than AIR for HetVeNet resource management, especially in large scale networks. Fig. 3 also shows that the cooperative communication can significantly improve FV rate compared to the no relay scheme.

Although the proposed MS-MAXMIN scheme archives lower total throughput of FVs than MS-

MAXSUN scheme as shown in Fig. 3, it can provide a significantly higher minimum VN rate compared to the others as plotted in Fig. 4, which means more fair resource sharing.

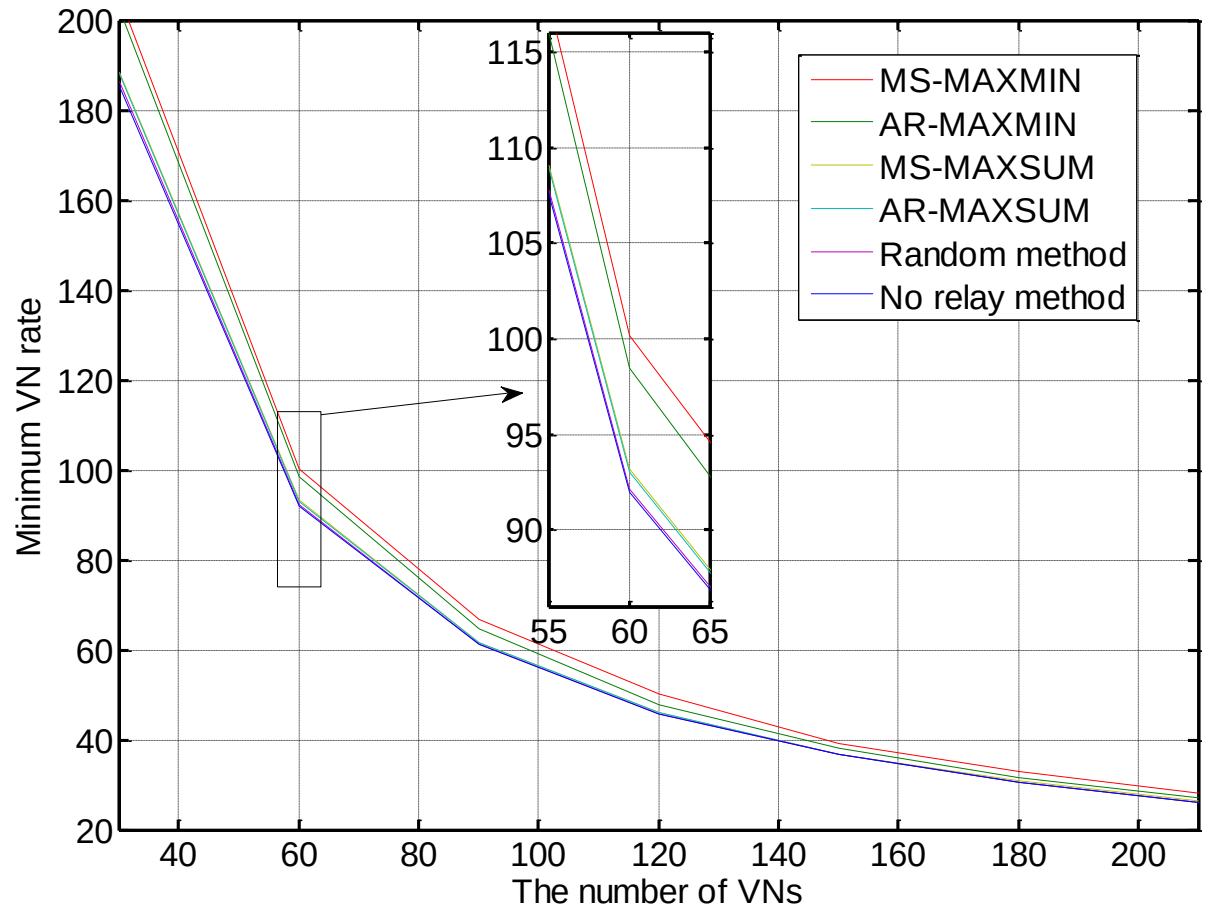

Fig. 4 Minimum VN rate vs. the number of VNs

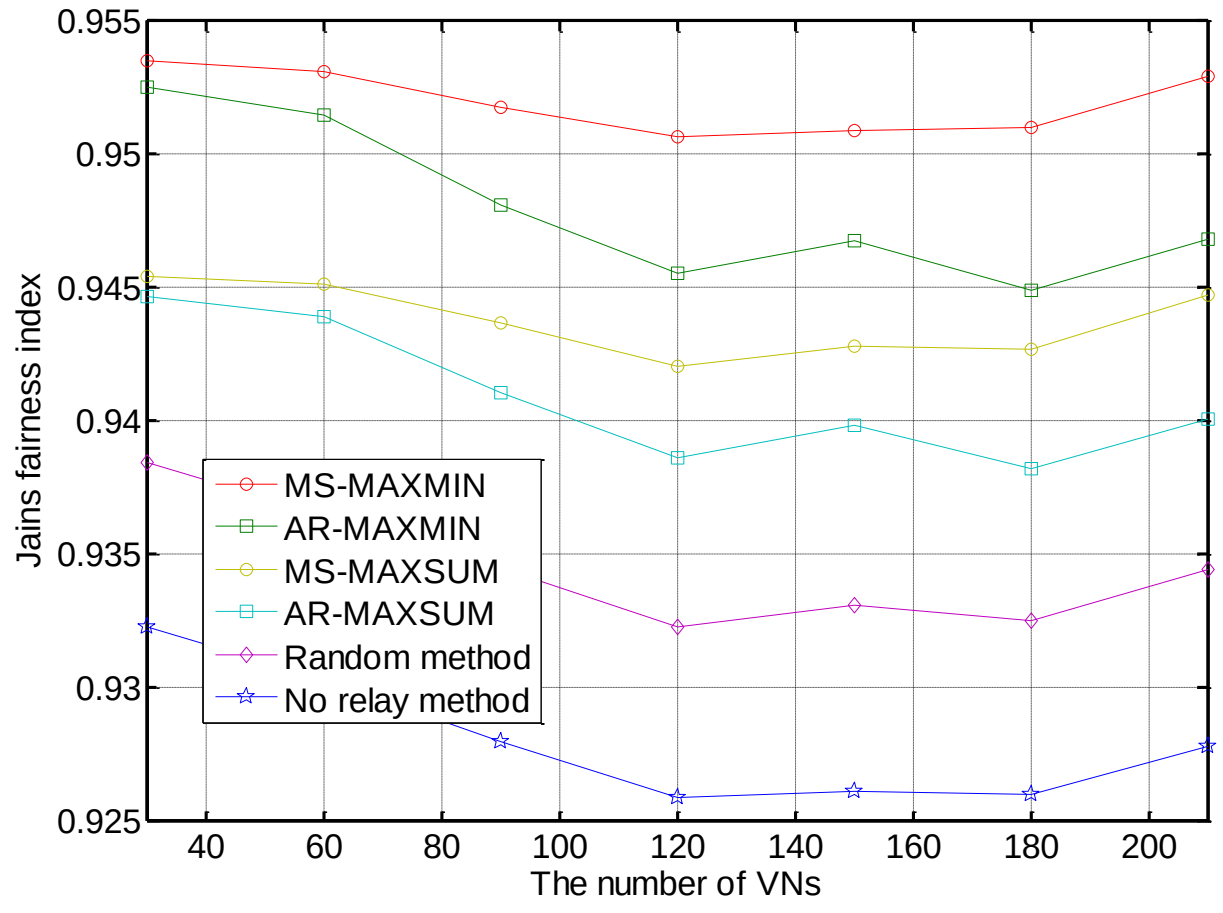

Fig.5 Jains fairness index versus the number of VNs.

The Jains fairness index [22] among VNs is defined as

$$FI(S_1, S_2, \cdots, S_N) = \frac{\left(\sum_{i=1}^{N} S_i\right)^2}{N \sum_{i=1}^{N} (S_i)^2},$$

which is a fairness measurement. The larger the $FI$ is, the fairer the allocation is ($1/N \leq FI \leq 1$). $FI=1$ when all VNs have the same allocation. Because the proposed scheme distributes resources fairly among HetVeNet members, the difference in the achievable data rates among the VNs can be decreased, which results in greater fairness. So the proposed MS-MAXMIN scheme can provide a significantly Jains fairness index compared to the others, as shown in Fig. 5.

## V. CONCLUSIONS

This paper studied the resource scheduling for heterogeneous vehicular networks. We proposed a mobile-service based max-min fairness resource scheduling scheme, where service amount is adopted to characterize the information transmission capacity of the links, which is more suitable for high mobility scenarios, and the max-min criteria is adopted to meet the fairness requirement among the moving vehicles. Simulation results demonstrated the effective of our proposed scheme and it's shown that our proposed scheme archives higher throughput compared with random scheduling scheme and no relay scheme and higher fairness compared with max-sum scheme, random scheduling scheme and non-relaying scheme.

## ACKNOWLEDGEMENTS

This work was supported by the Chinese Postdoctoral Science Foundation (no. 2015M570937), and the Open Research Fund of National Mobile Communications Research Laboratory, Southeast University (no. 2014D03).